\documentclass[preprint,aps,showpacs,showkeys,amsmath,amssymb,12pt]{revtex4}
\usepackage{graphicx}
\usepackage{dcolumn}
\usepackage{bm}
\usepackage{epsfig}
\usepackage{color}

\begin{document}


\title{
Quantum computing on long-lived donor states of Li in Si
}
\author{V. N. Smelyanskiy$^1$, A. G. Petukhov$^2$ and V. V. Osipov$^3$}

\affiliation{$^1$ NASA Ames Research Center, Moffett Field, CA 94035\\
$^2$ Physics Department, South Dakota School of Mines and
Technology, Rapid City, SD 57701\\
$^{3}$Hewlett-Packard Company Laboratories 1501 Page Mill Road,
1L-12, Palo Alto, CA 94304}

\begin{abstract}
We predict a gigantically long lifetime of the first excited state
of an interstitial lithium donor in silicon. The nature of this
effect roots in the anomalous level structure of the {\em 1s} Li
manifold under external stress. Namely, the coupling between the
lowest two states of the opposite parity is very weak and occurs
via intervalley phonon transitions only. We propose to use these
states under the controlled ac and dc stress to
process quantum information.  We find an unusual form of the
elastic-dipole interaction  between 
different donors. 
This interaction scales with the
inter-donor distance $R$ as $R^{-3}$ or $R^{-5}$ for the
transitions between the states of the same or opposite parity,
respectively.  The long-range $R^{-3}$ interaction provides  a
high fidelity mechanism for 2-qubit operations.

\end{abstract}

\maketitle

A broad effort is currently underway to develop new techniques and
systems for quantum information processing (QIP) that involves
coherent manipulation and control of a large array of 2-level
systems
(qubits) for quantum storage and computing purposes.
Implementation of the quantum computer (QC) as a solid-state device
has a number of unique advantages related to scalability and the
prospect of integration with state-of-the-art semiconductor device
fabrication technology. One of the main challenges in developing a
solid-state QC is the requirement of a relatively low qubit decoherence
rate. The natural candidates in this respect  are  QC schemes
based on spins as quantum bits, e.g.  nuclear spins of $^{31}$P
donors \cite{Kane}, 
or electron spins in semiconductor quantum dots \cite{Hetero}.
While spins are relatively well isolated from the
environment and characterized by long decoherence times, the
implementation of the spin-based QC is complicated by a short-range
character of exchange interaction. This provides additional
difficulties to very challenging problems of a single spin
measurement and control as well as \emph{nm}-precision fabrication
of a periodic qubit array. Additionally, the clock frequency for
nuclear spin QC is relatively low. The optically enhanced RKKY
mechanism for electron spin-exchange \cite{Sham:02} in
semiconductor quantum dots  allows for greater inter-qubit
distances, but it is not clear at  present how to make this QC
scheme scalable.

An interesting alternative to the spin-based qubits  is provided
by charge-based qubits  formed by the two lowest states of a
single-charged pair of quantum dots
\cite{Fedichkin:00,Hayashi:04}, or donor atoms
\cite{Hollenberg:04}. Those systems do not depend on the
single-spin readout and rely on a long-range electric-dipole
interaction between the qubits for performing quantum logic
operations. The  QC scheme proposed in  \cite{Golding:03} uses
acceptors in silicon to encode qubits with a decoherence rate
$\sim$ 1 KHz and relies on a long-range elastic-dipole interaction
between qubits ($\sim R^{-3}$). A serious source of decoherence in
all charge-based qubits is due to charge fluctuations in the
surrounding environment \cite{Paladino:02}. In particular,
relaxation of the charge traps can produce a shot noise signal
leading to decoherence times that are significantly shorter than
that in the spin-based QC
\cite{Hollenberg:04}. 

In this Letter we exploit the multivalley nature of the Si
conduction band and anomalous structure of the ground state of an
interstitial Li donor
to demonstrate that under certain conditions  the first excited
state of the single Li donor can have a lifetime comparable to
that of a nuclear spin (in access of 1 sec). The two lowest
\emph{1s} states of the Li donor in Si can be used to encode the
qubit that is extremely well isolated from the electromagnetic
fluctuations of the surrounding.

The conduction band of Si has six valley minima ${\bf k}_i$
located along six equivalent  directions, $[\hat x,-\hat x,\hat
y,-\hat y,\hat z,-\hat z]$  at about 85$\%$ of the distance
2$\pi/a_{\textrm{Si}}$ to the Brillouin zone boundary
 (here $a_{\textrm{Si}}$ is a lattice constant
for Si). In the framework of the effective mass theory (EMT)
the ground state of a shallow donor electron is 6-fold
degenerate. The
valley-orbit interaction removes the degeneracy and gives rise to
the splitting of the donor levels \cite{KL:55}. Excited states
within the ground-state manifold are produced by splitting of
the \emph{1s} hydrogenic energy levels and decay predominantly via
emission of acoustic phonons. The donor electron wave function can
be written in the standard form
$|\Psi^{\mu}\rangle=\sum_{|j|=1}^{3}\alpha_{j}^{\mu}\sum_{{\bf
k},{\bf G}} A^{j}_{{\bf k}+{\bf G}} |{\bf k}\rangle$. Here ${\bf
k}$ is a wave vector in the first Brillouin zone, ${\bf G}$ is a
reciprocal lattice vector,
 $ |{\bf k}\rangle$  are the Bloch functions and
index $\mu$ labels the irreducible representations of the point group
$T_d$ that are characterized by the coefficients
$\alpha_{j}^{\mu}$. Also,  $A^{j}_{{\bf k}}$ are the hydrogenic
envelope functions in $k-$space that are strongly localized  in
the vicinity of the corresponding valley center ${\bf  k}_j$ and
the summation above is taken over all valleys $j=\pm1,\pm2,\pm3$.

We now consider the acoustic phonon driven  transition rate  from a state
$|\Psi^{\mu'}\rangle$ to $|\Psi^{\mu}\rangle$:
$W_{\mu\mu^{\prime}}=2\pi/\hbar^2 \sum_{{\bf
q}\nu}|V_{\mu\mu^\prime}({\bf
q}\nu)|^2\delta(\omega_{\mu^{\prime}\mu}- \Omega_{{\bf q}\nu})$,
where $\hbar \omega_{\mu'\mu}$ is the change in the donor energy
during the transition, $\Omega_{{\bf q}\nu}$ is the frequency of
an acoustic phonon mode $\nu$ with the  wavevector ${\bf q}$, and
$V_{\mu\mu^\prime}({\bf q}\nu)$ are the matrix elements of the
electron-phonon interaction Hamiltonian, $\langle
\Psi^{\mu}|H_{\textrm{el-ph}}|\Psi^{\mu^{\prime}}\rangle=\sum_{{\bf
q}\nu}V_{\mu\mu^\prime}({\bf q}\nu) (b_{{\bf q}\nu}+b_{-{\bf
q}\nu}^{\dag})$, where
\begin{equation}
V^{\mu\mu^\prime}_{{\bf
q}\nu}=\sum_{|i|,|j|=1}^{3}\alpha_{i}^{\mu*}\alpha_{j}^{\mu^{\prime}}
\sum_{{\bf k}, {\bf G}} A^{i\,*}_{{\bf k}}A^{j}_{{\bf k}+{\bf
q}+{\bf G}} {\cal M}_{{\bf k}{\bf q}\nu}. \label{Vqnu}
\end{equation}
\noindent Here ${\cal M}_{{\bf k}{\bf q}\nu}$ is the
matrix element of $H_{\textrm{el-ph}}$ between the  Bloch
functions $|{\bf k}\rangle$ and $|{\bf k}+{\bf q}\rangle$
\cite{vogl1976a}.

For the long-wavelength acoustic phonons, with $q\ll |{\bf k}_i-{\bf
k}_j|$, the dominant contribution to (\ref{Vqnu}) comes from the
\emph{intravalley} terms with $i=j$.
The only other important terms
correspond to the {\em intervalley umklapp} processes between the valleys
$j$ and $-j$ centered at the band minima ${\bf k}_j$
and ${\bf k}_{-j}^\prime={\bf k}_{-j}+{\bf G}_j$,
where ${\bf k}_{-j}=-{\bf k}_{j}$,
${\bf G}_j=(4\pi/a_{Si}) \hat
k_{j}$, and  $\hat{k}_j={\bf k}_j/k_j$.
The points ${\bf k}_j$ and ${\bf k}_{-j}^\prime$
lie on the opposite sides of the Brillouin
zone boundary and correspond to the closest possible
intervalley separation $\kappa_0=0.6\pi/a_{Si}$.
All the other intervalley
($i\ne j$) and
umklapp terms (${\bf G}\neq 0$) are  much smaller, since they are
proportional to the overlap of the envelope functions
from much farther separated  valleys.
For both the intravalley and intervalley ($j\rightarrow -j$)
umklapp processes the sum  over ${\bf k}$ in Eq.~(\ref{Vqnu})
is dominated by the small vicinity of the
valley centers, where   ${\cal M}_{{\bf k}{\bf q}\nu}\approx {\cal
M}_{{\pm\bf k}_j{\bf q}\nu}\equiv M^{j}_{{\bf q}\nu}$, and
\begin{equation}
M^{j}_{{\bf q}\nu}= \left({\hbar}/{2\rho V \Omega_{{\bf
q}\nu}}\right)^{1/2}\left[\Xi_u (\hat k_{j}\cdot {\bf  q})(\hat
k_j\cdot \hat e_{{\bf q \nu}})+\Xi_d({\bf q}\cdot \hat e_{{\bf
q}\nu})\right]. \label{M}
\end{equation}\noindent
Here  $\Xi_u$ and $\Xi_d$
are the deformation potential constants, ${\hat e}_{{\bf q}\nu}$
is the polarization vector of a phonon mode ${\bf q}\nu$, $\rho$
and $V$ are the mass density and volume of the crystal,
respectively. Finally, we obtain
\begin{equation}
V^{\mu\mu^\prime}_{{\bf
q}\nu}=\sum_{|j|=1}^{3}\!\!\sum_{\bf k}
M^{j}_{{\bf q}\nu}\,
\alpha_{j}^{\mu *}A^{j*}_{\bf k}\left[
\alpha_{j}^{\mu^{\prime}}
 A^{j}_{\bf k+q}
+\alpha_{-j}^{\mu^{\prime}}
A^{-j}_{{\bf k+q+G}_j}\right],\label{ii}
\end{equation}
\noindent

The EMT donor eigenfunctions $\Psi^{\mu}({\bf r})$
have a certain parity which is equal to the parity of the multiples
$\alpha_{j}^{\mu}A_{\bf k}^{j}$ under the operation ${\bf
k}_j\rightarrow -{\bf k}_j$. For example, the parity is even for the
$s$-type singlet or doublet donor states belonging to the irreducible
representations $A_1$ or $E$, 
but it is odd for
$s$-type triplet states belonging to $T_2$ etc.
In the limit of small ${\bf q}$ the intervalley processes described
by the second term in Eq.~(\ref{ii})
give rather small contribution
and presumably can be neglected. Then, as it
follows from Eq.~(\ref{M}) and~
(\ref{ii}), the
matrix elements between the states of
the opposite parity, $V^{\mu\mu^\prime}_{{\bf
q}\nu}=0$, due to the cancellation of the intravalley  $j$ and $-j$ terms.
Based on this observation one can expect
that the probabilities of the even-odd transitions involving
long-wave acoustic phonons are much smaller than those of the
same parity transitions.
Analogous symmetry arguments explain a strong suppression of the
$1s(A_1)\rightarrow 1s(T_2)$ Raman transition in silicon donors
\cite{Klein:78}. One can also show that if the intervalley
processes  in (\ref{Vqnu}) are neglected the
transitions between the opposite-parity donor states
are forbidden in \emph{all orders} in $V^{\mu\mu^\prime}_{{\bf
q}\nu}$.

\begin{figure}[htb]
\includegraphics[width=.6\linewidth,angle=270
]{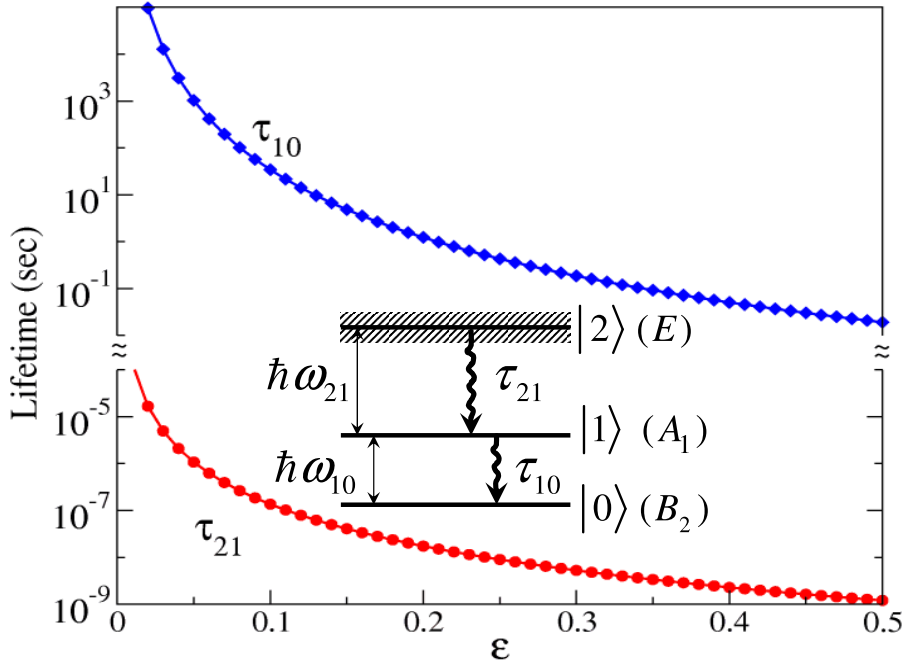} \caption{Dependence of the lifetimes $\tau_{10}$,
$\tau_{21}$ of the first two excited states of Li donor in Si on
$\varepsilon$. The insert  shows the energy levels of an isolated
Li donor in Si under the compressive force ${\bf F}\|\langle
001\rangle$. Bohr frequencies
$\omega_{10}=\omega_{21}/2=\varepsilon\Delta_c/\hbar$.}
\label{level_structure}
\end{figure}

We now consider implications of this effect for
the lithium donor in silicon. Li is
an interstitial impurity with the $T_d$-site symmetry. It has an
anomalous fivefold degenerate $1s(E+T_2)$ ground state while the
fully symmetric  state $1s(A_1)$ lies above, with the splitting
between the states conventionally denoted as $6\Delta_c=$1.76 meV
\cite{Aggarwal:65}. If a uniaxial compressive stress $F_z$ is
applied along the $\langle 001\rangle$ direction the site symmetry
is reduced from $T_d$ to $D_{2d}$ and has a distinct
mirror-rotation axis $\hat k_{3}$.  Then the ground state level is
split into three levels as shown in Fig.\ref{level_structure}
\cite{Aggarwal:65,Jagannath:81}.
 The new ground state denoted as $|0\rangle$ has
 an odd parity  with the coefficients
$\alpha^{(0)}_{j}=(0,0,0,0,1,-1)/\sqrt{2}$. The first excited
state, $|1\rangle$, is a singlet, it has  an even parity,
$\alpha^{(1)}_{j}=(b,b,b,b,a,a)$ where $a$= $a(\varepsilon)$,
$b$=$b(\varepsilon)$.
%
At small magnitude of the applied stress
$|F_z|$ one has \cite{Jagannath:81}: $a=(6+\varepsilon)/6\sqrt{3}$,
$b=(3-\varepsilon)/6\sqrt{3}$, where
\begin{equation}
\varepsilon=\Xi_{u}(s_{11}-s_{12})F_z/{3\Delta_c},
\label{epsilon}
\end{equation}
and $s_{11}$, $s_{12}$ are elastic compliance constants. The
second excited  energy level  is a triplet, it  consists
 of an even parity state,  $|2\rangle$,
with $\alpha^{(2)}_{j}=\frac{1}{2}(1,1,-1,-1,0,0)$ and two  odd
parity states. 

From the above it follows that only the  second term in Eq.~(\ref{ii})
with $i=-j=\pm 3$
contributes to the phonon decay rate $W_{10}$ of the  state $|1\rangle$.
%
The decay rate 
calculated with
the Kohn-Luttinger envelope functions $A_{\bf k}^j$ \cite{KL:55}
reads
\begin{equation}
W_{10}\equiv \tau_{10}^{-1}=\frac{2}{35}
\cdot\frac{a^2(\epsilon)(a_\parallel\kappa_0)^2}
{[1+(a_\parallel\kappa_0/2)^{2}]^6}
\cdot\frac{\Xi_u^2\omega_{10}^5a_\parallel^2}{\pi\hbar\rho u_t^7}.
\label{W10}
\end{equation}
\noindent Here  $u_t$ is the transverse speed of sound, $a_{\|}$ is
the
longitudinal Bohr radius for the  \emph{1s} donor state in Si,
and we assumed that  $2\pi u_t/\omega_{10}\gg a_{\|}$.
We  emphasize a very fast decrease of the transition rate
 between the opposite parity states
with the level separation, $W_{10}\propto \omega^{5}_{01}$. This
is in contrast with the conventional dependence of the transition
rate between the same parity states
\cite{Golding:03,Fedichkin:04}. For instance,
$W_{21}\propto
\omega_{21}^{3}$  for the transition
$|2\rangle \rightarrow |1\rangle$. We note that the decay rate $\propto \omega^5$
was found in \cite{Fedichkin:04} for the electron transition in a
singly-charged double-dot system between  the antisymmetric excited
state and symmetric ground state separated by the energy gap
$\hbar \omega$. However, a key distinction of the expression
(\ref{W10}) from the results of Ref.~\cite{Fedichkin:04} is that it
contains an additional small factor, $8 a^{2}(\varepsilon)/35
(a_{\|}\kappa_0/2)^{10}\approx \textrm{10}^{-\textrm{5}}$,  due to
the overlap of the envelope functions from different valleys. It
is the {\em combination} of these two small factors in (\ref{W10})
that
gives rise to the giant lifetime of the state $|1\rangle$
at the low stress values (see Fig.
\ref{level_structure}).  This is one of the central results of this
work.

In addition to the energy decay 
there also exist the
dephasing processes \cite{Golding:03}  due to quasi-elastic phonon
scattering off donors. Detailed analysis shows that these
are not important at low temperatures. For example,
for $H_{\textrm{el-ph}}$  given by Eq.~(\ref{ii})
the dephasing rate $\tilde W_{10}=\nu_0 (T/T_{0})^{11}$, with
$\nu_0\sim 2\cdot 10^{14}$ Hz and $T_0=\hbar u_t/a_\bot$=19 K.
Thus at $T$=0.1 K
the dephasing rate is negligibly small. Finally, the dipole moment of
$|0\rangle\,$-$|1\rangle$ transition is less then
$\textrm{10}^{-\textrm{3}} e$\AA\, and therefore its quantum phase
is very robust against the charge noise.

The fact that electronic  states of a Si:Li donor  are extremely
well-isolated from the environment provides an interesting
possibility to consider this system for quantum information
processing with a pair of
states $|0\rangle$,  $|1\rangle$ 
used to encode a qubit. 
In what follows we make
use of the fact that  by varying the local pressure $F_z$ on a
given qubit one can
 change the frequencies $\omega_{\mu\mu'}$ and
therefore lifetimes $\tau_{\mu\mu'}$ of the corresponding
transitions. As seen in Fig.~\ref{level_structure} there is a
dramatic difference between the  lifetimes of the even-odd and
same-parity transitions  which exceeds seven orders of magnitude at
$\varepsilon=0.5$. For smaller $\varepsilon$  the lifetime
$\tau_{10}$  becomes extremely large, it reaches seconds and even
minutes for $\varepsilon\lesssim 0.2$ or $\hbar
\omega_{01}\lesssim$ 0.06 meV. This energy scale can be used
for a qubit level separation.

\begin{figure}[htb]
\includegraphics[width=.8\linewidth,
]{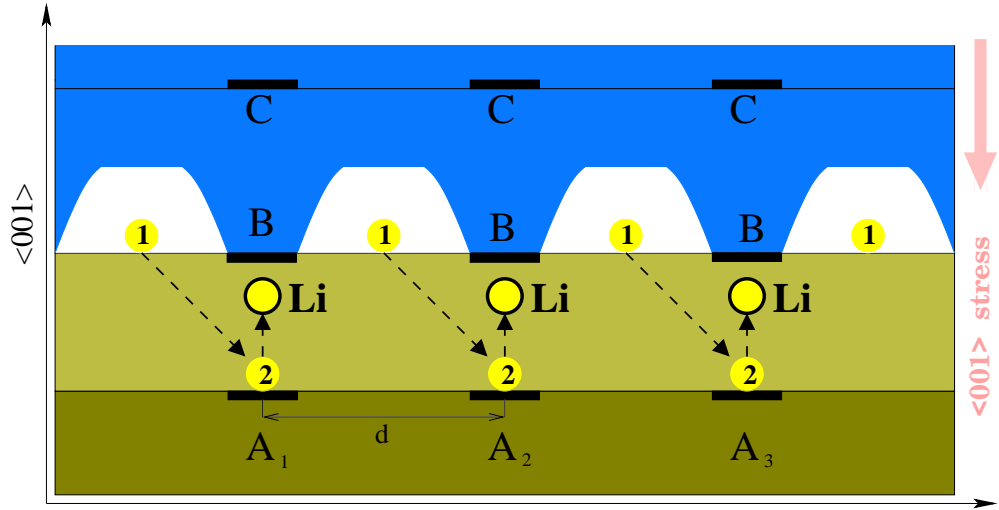}
\caption{
{\protect{Schematic of the Si:Li-based quantum
computer.  An etched
piezoceramic structure (blue), which is placed on the top Si layer (green),
has a row of columns  of a width
$w\sim$ 20 nm separated by arches.
The columns are used to generate a local $\langle
001\rangle$ pressure, controlled by the voltage biases between the
pairs of $B$-$C$ electrodes. The decay factor of the local stress
(and electric) field over the interqubit distance $d\sim$ 100 nm
is $\sim(w/2d)^2=10^{-2}$
\cite{Landau}. The residual  stress and electric field at
each Li donor arising from the
local control of the neighboring qubits can be compensated by
the synchronized  adjustment of voltages for all qubits.}} }
\label{layout}
\end{figure}

{One-qubit operations  that are required for
quantum computation with Si:Li donors can be implemented by
applying tailored pulses of stress or electric field {\em locally}
to selected donor atoms
(Fig.~\ref{layout}). 
For example, a
{\em rectangular stress pulse} of duration $\tau_{\rm dc}$  in
$\langle 001\rangle$ direction  will shift donor transition
frequency $\omega_{10}$ and produce a phase gate $\exp[i\tau_{\rm
dc}\Delta\omega_{10}\hat S_{z}]$. Here $\hat S_{z}$ is
a $z$-component of the
pseudo-spin-$\frac{1}{2}$ operator $\hat {\bf S}$ acting upon the qubit states
$|0\rangle\equiv |\hspace{-0.06in}\downarrow\rangle$ and
$|1\rangle\equiv |\hspace{-0.06in}\uparrow\rangle$. One can also
apply a low-amplitude
{\em periodically modulated stress pulse}  $A_{\langle
001\rangle}\cos\omega_{10} t$ of duration $\tau_{\rm ac}$  in $\langle 001\rangle$ direction.
This pulse
will resonantly drive the transition $|0\rangle\rightarrow|1\rangle$ of a
given qubit. It will produce a pseudo-spin rotation around
 $x$-axis, $\hat X(\varphi)\equiv \exp[i \varphi\hat S_{x}/2]$,
 $\varphi\equiv 2\tau_{1}\Omega_x$. 
 An estimate for $\Omega_{x}$
 in the rotating wave approximation gives}
\begin{equation}
\Omega_{x}=\frac{128 \,A_{\langle 001\rangle}\omega_{10}
s_{11}}{u_{l}\sqrt{6}}(\Xi_u+\Xi_d) \kappa_0
a_{\|}^{2}.\label{Omega}
\end{equation}
\noindent
{Here $u_l$ is the longitudinal speed of
sound). 
With a modest ac stress amplitude $A_{\langle 001\rangle}\simeq$
10$^{\rm 5}$ dyn/cm$^{\rm 2}$ and a
 driving frequency $\omega_{10}/2\pi$=10 GHz one gets
$\Omega_{x}/2\pi\simeq$630 MHz. Then the $\pi$-pulse $\hat X(\pi)$
will have a duration $\tau_{1}\simeq$ 0.4 ns and a corresponding
decoherence-induced error  $\sim \tau_{1}W_{10}$ will be
less then  $10^{-9}$.}

{Another controlling option consists in applying pulses of the
{\em time-dependent electric field} ${\bf
E}(t)\|\langle001\rangle$  caused by the modulation of a bias voltage
between the local electrodes ($A,B$ in Fig.~\ref{layout}). The electric field will
produce a quadratic Stark shift of
$\omega_{10}$ and will induce the Rabi oscillations between the
qubit states $|0\rangle$ and $|1\rangle$ via the virtual transitions
through the $2p_0$ state. For an electric field amplitude $E\sim
\textrm{10}^\textrm{3}$ V/cm the period of the Rabi oscillations
$\tau_{1} \propto E^{-2}$ is of the order of 0.1 ns.}

Li donors in Si are coupled
to each other via acoustic phonon field and behave as
elastic dipoles. As in the  case of
the shallow acceptors in Si \cite{Golding:03}, this coupling can be
used as a \lq\lq data bus" to implement two-qubit logic gates.
The matrix elements of the phonon-mediated coupling Hamiltonian,
$H_{ij}=\sum |\mu_i\mu_j\rangle G^{ij}_{\mu_i\mu_i^\prime,
\mu_j\mu_j^\prime}\langle\mu_i^\prime\mu_j^\prime| + h.c.$,
between a pair of
donors at the sites ${\bf R}_i$ and ${\bf R}_j$ can
be  expressed as follows:
\begin{equation}
G^{ij}_{\mu_i\mu^{\prime}_{i},\mu_j\mu^{\prime}_{j}}
=\frac{1}{\hbar}\,\sum_{{\bf q}\nu}
V^{\mu_i\mu^{\prime}_{i}}_{{\bf q}\nu}
\left(V^{\mu_j\mu^{\prime}_{j}}_{{\bf q}\nu}\right)^{*}
\frac{e^{i{\bf q}({\bf R}_{i}-{\bf R}_{j})}}{\Omega_{{\bf q}\nu}},
\label{rabi}
\end{equation}
\noindent where
we assumed that
$\omega_{\mu\mu^{\prime}}|{\bf R}_i-{\bf R}_j|/u_t\ll 1$.

Consider first a
resonant excitation transfer (RET) between  donors \emph{i} and
\emph{j} involving a pair of transitions, $\mu\rightarrow\mu^\prime$
on site \emph{i} and $\mu^\prime\rightarrow\mu$
on site \emph{j}, that are in resonance with each other.
The coupling constant corresponding to this process
$\hbar g^{\mu\mu^\prime}_{ij}=G^{ij}_{\mu\mu^{\prime},\,\mu^\prime\mu}$.
For the RET involving the even-odd transitions
$|0\rangle\leftrightarrow|1\rangle$
this
coupling constant
reads
\begin{equation}
\frac{g^{10}_{ij}}{W_{10}}=\frac{315}{16}\left(3-\frac{u_t^2}{u_l^2}\cdot
(4\sigma+5) \right)\cdot\left(\frac{u_t}{\omega_{10}R_{ij}}\right)^5.
\label{g10}
\end{equation}
Here $\sigma=\Xi_d/\Xi_u$ and we assumed that ${\bf R}_{ij}=
{\bf R}_{i}-{\bf R}_{j}$
lies in the plane normal to  $\langle 001\rangle$.
We note an unusually steep falloff of the coupling
constant with the donor-donor separation.  In fact, one can show that
if the RET involves transitions between the states of the same parity,
a conventional dependence $g_{ij}\propto 1/R_{ij}^3$
\cite{Golding:03} is recovered. In particular, the RET coupling
constant for the $|1\rangle\leftrightarrow|2\rangle$ transition in
Si:Li is
\begin{equation}
\label{g21}
g^{21}_{ij}/W_{21}=\gamma\,(u_t/\omega_{21}R_{ij})^3
\end{equation}
where $\gamma=(5/16)(2+7(1-u^2_t/u^2_l))\simeq$2.0.
This long-range dipole interaction is due
to the intravalley terms in
 $V_{{\bf q}\nu}^{21}$ that are absent in $V_{{\bf q}\nu}^{10}$.
It can be used to selectively
generate an entanglement between two donors
by tuning their energy levels into resonance with each other.

Consider now a diagonal part of the phonon-mediated donor-donor
Hamiltonian
generated by the diagonal matrix elements $V^{\mu\mu}_{\bf{q}\nu}$
in Eq.(\ref{rabi}). This
is an Ising-type dipole-dipole interaction resulting  from the
dependence of  a polaron shift in the electron energy of a given
donor on the electron states of the  neighboring donors. The
effective interaction Hamiltonian projected onto the states $\mu=0,1$ of
$N$ donors located at
 ${\bf R}_j$  has the
form:
\begin{equation}
{\cal H}(t)=
\frac{\hbar}{2}\sum_{i\ne j=1}^{N} \left[J_{ij}
\hat S_{iz}\,\hat S_{jz}+ g_{ij}
\hat S^+_i\hat S^{-}_j + h.c.\right]
+
\sum_{j=1}^{N}H_j(t).\label{H}
\end{equation}\noindent
{Here  $\hat S_{j}^+$ and  $\hat S_{j}^-$  are raising and lowering
 pseudo-spin operators for $j$th
qubit, $H_j(t)$  is a  single-qubit Hamiltonian,
$g_{ij}\equiv g^{10}_{ij}$,
and the Ising-exchange coupling
constant $\hbar J_{ij}=G^{ij}_{11,11}+G^{ij}_{00,00}-2G^{ij}_{00,11}$.}
Due to the
symmetry of the problem, $J_{ij}$ does not depend on the
in-plane orientation of ${\bf R}_{ij}\bot\langle 001\rangle$. Using
(\ref{M}),
(\ref{ii}),(\ref{rabi}) and the form of the valley-orbit
coefficients  one can obtain
\begin{equation} J_{ij}=\frac{\Xi_{u}^{2}}{32\pi\hbar\rho
u_t^2 R_{ij}^3}
\left(-1+\frac{5}{3}\cdot\frac{u_t^2}{u_{l}^{2}}\right).\label{J}
\end{equation}
\noindent Here the  $1/R_{ij}^3$ law  is also due to the intravalley terms
that are always present in the diagonal matrix elements $V_{{\bf
q}\nu}^{\mu\mu}$.  For $R_{ij}=$ 100nm, $J_{ij}/\pi \simeq$ 10 MHz while
$g_{ij}/\pi$=0.4 Hz. Because of the detuning  of  the donor
transitions from resonance the RET frequency at $|0\rangle\rightarrow|1\rangle$
transition
is further reduced and can always be neglected. Contrary to that, the
dipole interaction in ${\cal H}(t)$ always persists. It is a basic
long-range 2-qubit interaction in the liquid-state NMR  and also in
the solid-state QC scheme based on magnetic-dipole coupling
of donor spins in silicon \cite{Sousa:04}. The Hamiltonian~(\ref{H})
with $g_{ij}=0$ is sufficient to execute {\em any} one- or two-qubit
gates \cite{Sousa:04}. It serves as a basis of our QC scheme.

The QC scheme utilizing Ising-type dipole coupling has been
described  in Ref.~\cite{Sousa:04}.
It is based on the refocusing technique where an auxiliary set of
$\pi$ pulses is applied to a group of qubits $j=1:n$.  The
$\pi$-pulses cancel the dipole interaction terms in (\ref{H}) for
all members of this group except  for a selected pair of
neighboring qubits. The error of two-qubit gates, caused by the
non-exact cancellation, $p\sim(\tau_{1}/\tau_{2})^6$
\cite{Sousa:04}, where $\tau_1$ and $\tau_2=\pi/J_{j,j+1}$
are characteristic time scales for one- and two-qubit gates,
respectively.  For $R_{j,j+1}=100$ nm $\tau_2=10^{-7}$s and
$\tau_1 \lesssim$10 ns one has $p< 10^{-\textrm{6}}$. The error of
two-qubit gates due to the phonon-decay of the state $|1\rangle$
is extremely small. The corresponding quality factor $q$ for the
two-qubit gate
\begin{equation}
q\equiv \frac{1}{\tau_{2} W_{10}}=
\frac{35\left[1+(a_{\|}\kappa_{0}/2)^2\right]^6\,u_{t}^{5}\left(-1+5/3
\left(u_t/u_l\right)^2\right)}{ 64\pi
a^{2}(\varepsilon)\,(a_{\|}\kappa_{0})^2\,
R_{j\,j+1}^{3}\omega_{10}^{5}a_{\|}^{2}},
\end{equation}
\noindent is probably greater then $\textrm{10}^{\textrm{7}}$ at
$\omega_{\textrm{10}}$=0.06 meV/$\hbar$.

We note that a  frequency of resonant ac stress modulation
$\omega\simeq \omega_{10}$ is typically limited by a
high-frequency cutoff of the response function of a nano-scale
piezoelectric transducer (cf. Fig.~\ref{layout}). One can show
that for practically feasible cases this implies
$\omega_{10}\lesssim 10^{\rm 10}$ rad/s. However one could keep
the qubit energy difference at much higher levels, $
\overline{\omega}_{\rm 10}\simeq {\rm 10}^{\rm 11}$rad/s (0.1-0.06
meV), at all times, except the short time intervals when a given
qubit  is involved in a gate operation. Then the energy difference
can be adiabatically reduced the smaller value
$\omega_{10}\lesssim 10^{\rm 10}$ rad/s. If the time between gates
involving a given qubit is $\gg W_{10}^{-1}$ the qubit energy
levels will not be thermally repopulated and the working
temperature of a quantum computer can be set to $T^*\simeq \hbar
\overline{\omega}_{10}/k_B\simeq$ 50-100 mK.

Similarly, one can use the adiabatic stress reduction
in the alternative three-level QC scheme where a long-range RET,
based on the donor transition $1-2$  (see above) with Rabi
frequency $g^{21}_{ij}$ (Eq.~(\ref{g21})), is used to execute
2-qubit gates.
This process provides the most
direct tool to generate  entangled states $c
|1_{j}0_{j+1}\rangle+c'|0_{j}1_{j+1}\rangle$ between a given pair
of neighboring donors ($j,j+1$) whose
energy levels  are brought into to
resonance using locally applied stress pulses.
To achieve high-fidelity 2-qubit
gates within this scheme we need much smaller interlevel separation
$\hbar\omega_{21}$ than $\hbar\omega_{21}=$0.12~meV at
$\varepsilon=$0.2. Therefore
whenever
we need to execute a 2-qubit gate  we will adiabatically reduce the
stress $\varepsilon(t)$  from $\varepsilon_0=0.2$ to
$\varepsilon_2\simeq$ 0.002 ($F_z \sim \textrm{1.3}\cdot
\textrm{10}^{\textrm{5}}~\textrm{dyn}/\textrm{cm}^{\textrm{2}}$)
which will set the transition frequency to $\omega_{21}/2\pi
\simeq 0.24$~GHz.

\begin{figure}[tb]
\includegraphics[width=.7\linewidth,
]{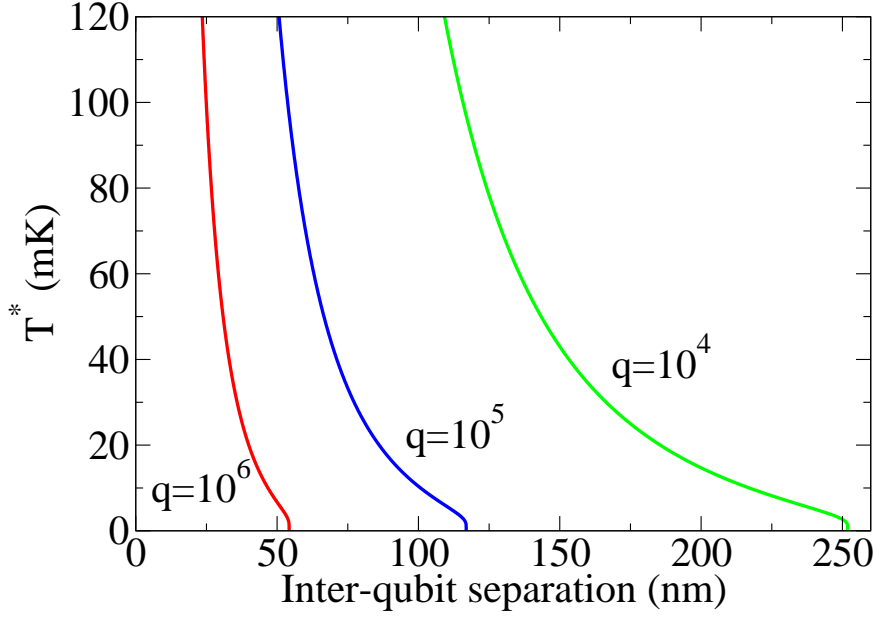} \caption{Quality factor for the 3-level quantum
computing scheme: dependence of the operational temperature $T^*$
on the inter-qubit distance $R$ for different values of the
quality factor $q$ and $F_z=\textrm{1.3} \cdot
\textrm{10}^\textrm{5} \textrm{dyn}/\textrm{cm}^\textrm{2}$
($\hbar\omega_{12}=0.001$ meV)}\label{temperature}
\end{figure}

Within the three-level scheme the Rabi frequency $g^{21}_{i,i+1}$
defines the QC clock frequency. In analogy to the above we define
the quality factor $q$ as the ratio of $g_{i,i+1}^{21}$ to the
phonon decay rate of the state $|2\rangle$ at the finite
temperature $T^*$:
\begin{equation}
\label{q}
q=\frac{g_{i,i+1}^{21}}{\left[n_{21}(T^*)+1\right]W_{21}}=
\frac{\gamma}{n_{21}(T^*)+1}\left(\frac{u_t}{\omega_{21}R}\right)^3,
\end{equation}
where $R\equiv R_{i,i+1}$ and $n_{21}(T^*)=
\left[\exp(\hbar\omega_{21}/k_BT^*)-1\right]^{-1}$ is the Planck's
distribution function. Consider a particular example with $R$=50~nm,
$\varepsilon=2\cdot 10^{-3}$, and $T^*=$100~mK.
 In this case the
 characteristic time  for a  2-qubit gate, $2\pi/g_{12}\simeq$0.2 $\mu$sec, is much
shorter than the relaxation time $\tau_{21}\simeq$ 3 msec and
therefore the donor electron levels will not be thermally
repopulated during RET despite $k_B T^*/\hbar\omega_{12}\simeq$ 8.
As soon as the two-qubit gate is executed the qubits will be
adiabatically returned into their high-stress state. With this
choice of parameters the QC clock frequency is
$g_{21}/2\pi\simeq5.2$~MHz and the quality factor of a 2-qubit
gate $g_{21}/W_{21}\sim 10^5$. Solving Eq.~(\ref{q}) for $T^*$ we
obtain the operational temperature as a function of desired
quality factor $q$ and inter-qubit separation $R$:
\begin{equation}
\label{T}
T^*=-{\hbar\omega_{21}}/{3k_B\ln\left(1-\frac{q}{3\gamma}
\cdot\frac{\omega_{21}R}{u_t}\right)}.
\end{equation}
By lowering the operational temperature $T^*<$ 100 mK we can
significantly increase the quality factor or/and increase the interqubit
separation $R$ as shown in
Fig.~\ref{temperature}.

By varying the locally applied stress $F_z$ and  voltage
on the local electrodes (see Fig.~\ref{layout}) one can
selectively shift the $1s(E+T_2)-2p_0$ transition frequencies of a
given donor atom by $\hbar \Delta\omega\sim$ 0.3 meV without
significantly affecting the qubit level separation $\hbar
\omega_{10}$. Then the donor excitation lines $1s(E+T_2)\rightarrow 2p_0$ can
be selectively brought into resonance with a {\it globally}
applied infrared radiation field since their homogeneous broadening
($\simeq$55 MHz \cite{Orlova:00}) is $\sim$500 times smaller
then the frequency shift $\Delta\omega$. Such a selective resonant
excitation can be used to perform a measurement of the qubit state.
Indeed, according to the polarization selection rules
\cite{Jagannath:81} a linearly polarized electric field ${\bf
E}\bot{\bf F}$ will resonantly excite the transitions from the
state $|1\rangle$ to the states of the $2p_0$ manifold but will
{\bf not} affect the donor electron that was in the state
$|0\rangle$ before the IR pulse arrived.
Detection of the dipole moment
of this state
($\sim$ 40~\AA$e$) by means of a single electron transistor
(SET) measurement
will determine whether the state $|1\rangle$ was occupied prior to
the excitation.
The fabrication of the Si:Li QC will utilize  very large
electro-mobilities  of Li ions in Si
(at
$T$=400K and $E$=${\rm 10}^{\rm 6}$V/cm the drift velocity  is 0.4
cm/s). By controlling 
the electric field
between the electrodes $A$ and $B$ (Fig.~\ref{layout}) the Li ions,
sequentially deposited on the surface of a super-pure Si layer
(positions 1), will
be dragged to their target positions 2 to form a subsurface
array
with the period $d\simeq $ 100nm.
Subsequent lowering of the temperature to its
operational value $T\sim $ 50-100 mK will form a stable regular
array of the neutral Li donors.

In conclusion, we have shown that the valley-orbit
structure of the ground and first excited states of the Si:Li donor
gives rise to a gigantic lifetime of the excited state.  We
have found that the elastic dipole
coupling between two donors scales with the
inter-donor distance $R$ as either
$1/R^3$ or $1/R^5$, depending on their
electronic states.
We build on these findings
and propose a QC scheme with the spinless stress-defined elastically
coupled qubits. 
It is characterized by extremely small error rate (lesser then
$10^{-6}$), relatively large operational temperatures and
interqubit separations that are 5 to 10 times greater then
those used in the nuclear spin solid state QC.

We acknowledge  valuable discussions with M. Dykman, B. Golding,
and M. Foygel. This work was supported in part by  NASA
Revolutionary Computing Program (V. S.) and NSF (A. P.).

\end{document}